\documentclass[12pt,preprint]{aastex}

\slugcomment{Article for DM 2004 - Edinburgh}

\shorttitle{Baryonic Dark Matter Consequences}
\shortauthors{ Schild}

\begin{document}

\title{Some Consequences of the Baryonic Dark Matter Population}

\author{Rudolph E. Schild}
\affil{Center for Astrophysics,
        60 Garden Street, Cambridge, MA 02138}
\email{rschild@cfa.harvard.edu}

\begin{abstract} 

Microlensed double-image quasars have sent a consistent
message that the baryonic dark matter consists of a dark population of
free-roaming planet mass objects. This population has long been predicted
to have formed at the time of recombination, 300,000 years after the Big
Bang, when the primordial plasma changed to neutral atoms with an attendant
large increase in viscosity of the primordial matter. Following a very
brief review of the observational basis for this conclusion and some
alternative explanations, we review some probable effects of this population.
After the particles formed by the usual gravitational condensation - void
separation process, they collapsed on a 100 million year Kelvin-Helmholz
time scale, and started their inevitable cooling process. Although not yet
satisfactorily modeled, this process should have caused significant
evaporation of primordial gas and taken them through the condensation and
freezing points of hydrogen on their way to the 2.73 K temperature of the
present universe. At the 20 K freezing point they should have frozen from
the outside in, creating tremendous crushing  central pressures that 
would have
easily produced the rocky cores of planets and Kuiper-Oort cloud objects
mysteriously over-abundant in the present solar system. 
The mystery of how did the universe become re-ionized by a Pop III that
should have been seen at redshifts 6 to 8, now under scrutiny from direct
spectroscopic observation, is cleanly side-stepped. Probably $99\%$ of
the baryonic matter in the universe was sequestered away in 
the dark matter bodies
and does not need to be re-ionized for the universe to have its present
transparency in the far ultraviolet.
And the Dark Energy mystery will evaporate when it is understood how this
population reduces the transparency of the universe. It is probably not a
coincidence that the "self replenishing dust" model that explains the HST
supernova brightness deficits closely matches the known dependence of
extinction from $Ly-\alpha$ clouds upon redshift. If these mysterious clouds,
that should have diffused away on a short time scale, are reforming from
slow evaporation of the planet-mass population, they should produce
spherical lenses that refract light out of the supernova images to produce
a grey reduction of the transparency of the universe.

\end{abstract}


\section{Introduction and Microlensing Results in Q0957}

At the 1996 Sheffield Symposium I reported microlensing 
results from 15 years of Q0957 monitoring that revealed a persistent,
continuous rapid microlensing which indicated that compact planetary mass
bodies constitute the baryonic dark matter (Schild, 1996). 
At the time, the response was,
appropriately, "That's nice, but it's just Rudy's data for Rudy's quasar."
Today, 8 years on, the rapid microlensing has been confirmed in the 6
additional lens systems having sufficient data for time delay estimation.
And new Q0957 observing campaigns have produced evidence for an event with
amplitude $1 \%$ and time scale of just 12 hours (Colley \& Schild 2003).
A recent summary of these observational results was presented
at the UCLA Dark Matter/Dark Energy 2004 Symposium (Schild, 2004;
astro-ph/0406491).

With the observation of rapid microlensing now confirmed, alternative
explanations based upon imagined possible quasar structure have fortunately
been explored. Thus bright points orbiting in the quasar's accretion disc
have been considered (Gould and Escude-Miralde, 1997; 
Schechter {\it et al} 2003)
and dark clouds swarming around the quasar (Wyithe $\&$ Loeb, 2002)
have been investigated. These
schemes have been frustrated by the extreme physics required, especially
for the extremely rapid Colley \& Schild (2003) event, and by the implied
periodicity. But most important, the simulations do not naturally produce the
equal positive and negative events seen in the Schild (1999) wavelet
analysis and natural to the planet-mass microlensing explanation 
(Schild \& Vakulik, 2004).
These points were more extensively reviewed in Schild (2004).

The above results are based upon reasonably rigorous science, with
simulations and confirmed observational results. For the remainder of this 
contribution I examine more tentative explorations, based largely upon
back-of-the-envelope calculations, to examine some immediate applications
of the discovery. For example, what would such particles look, feel, smell,
and taste like?

\section{What do the Particles Look, Feel, Smell, and Taste Like?}

Remarkably, these questions were already answered at the 1996 Sheffield
symposium by hydrodynamics expert Carl Gibson, who predicted their 
existence from hydrodynamical analysis of the forces operative in the
fluid/gas of the expanding universe at times before, during, and after
recombination, 300,000 years after the Big Bang. However this
hydrodynamical theory has been largely ignored because of its departures
from the prevailing CDM theory, whose failures will need to go to
completion before the hydrodynamical theory features receive the scrutiny
that they deserve.

In the Gibson (1996, 2000) theory, structure formation limited by
viscous-diffusive forces already seeded structures on scales
of galaxies, clusters, and superclusters during the
plasma epoch, before recombination. After the plasma-gas transition
the baryonic gas of the universe fragmented at Jeans mass clumps 
further seeded with
planetary mass "Primordial Fog Particles" (PFP's) limited to sub-stellar
mass scales by the viscous and gravitational forces in the fluid. Thus
the entire baryonic mass of the universe gravitationally collapsed onto 
primordial scale-free fluctuations but only the planet mass ones succeeded;
they survive today as the baryonic dark matter seen in quasar microlensing.
Many or most of these PFP particles would in turn be initially contained in
globular cluster scale condensations that are often seen today mysteriously
appearing in galaxy interactions.

The PFP's presumably collapsed on the 100 million year Kelvin-Helmholz time
scale and then cooled. In their subsequent history they would have
swept up dust from supernovae and cool giant stars. In the expanding 
universe they would presumably have cooled below the hydrogen
condensation point, and when passing below the freezing point at 20 K they
would have crushed the ices and rocks at their centers to make the solid
cores of the planets and the Trans-Neptunian objects seen today.

Gibson (1997) described the PFP's today as "in solid or liquid state,
crusted with 14 billion years of accreted dust." It also seems likely that
these are the objects at the cores of the Lyman-alpha clouds seen by the
thousand in ultraviolet spectra of quasars. 

Today, we do not have an adequate simple model of a planetary mass particle
formed at recombination and passively collapsing, reaching its maximum
temperature at around 100 million years after recombination, and subsequently
cooling. We don't even know if the PFP's would be cool enough at their
centers for hydrogen to cool through the hydrogen condensation point at 40
K and the freezing point at 20 K. Because
galaxy discs are now measured to have temperatures of 30 K 
(Bendo {\it et al}, 2003), there should
also have been a point in the asymptotic approach to our 2.73 K background
temperature when such Halo PFP objects orbiting galaxy discs went through
multiple freezing and melting cycles.

Because the PFP objects are dark, they would be detected by 
my gravitational microlensing or by 
chance superpositions against a bright nebular background, as
probably seen in the nearest planetary nebula, the Helix (O'Dell \&
Handron 1996). There they have been
interpreted as hydrodynamical or shock instabilities in the
expanding shell, even though such interpretations are contradicted by the high
masses and densities measured for them (Meaburn {\it et al}, 1998). 
Interpreted as PFP's seen against
the bright nebular background, with surface ablation wearing away their
outermost layers, their masses are estimated as $10 ^{-5} M_\odot$ and their 
diameters are measured to be $10^{16}$ cm, or 100 times the diameter of
Neptune. 

These Helix cometary knots are seen in the HST images to occur mostly in
clumps, as expected for self-gravitating particles. This would explain why
they could not be detected by MACHO microlensing searches, where they would
just cause rapid irregular variability rejected by the MACHO search software.

\section{The Chemical Enrichment and Re-Ionization of the Universe.}

Our cosmology today is challenged by the observational result that
reionization of the universe occurred at redshifts around 6.5, indicated by
the ultraviolet continuum of hi-z quasars (Fan {\it et al}, 2004). 
Thus a vast population
of high-redshift galaxies rich in Pop III stars
should exist and supply copious  $Ly-\alpha$ photons
causing the universal ionization. But actual observations of galaxies at
the redshift range 6.5 - 7.5 do not show copious $Ly-\alpha$ emission or
extreme luminosity; indeed, the searches produced the conclusion that
galaxies are x9 less luminous at the highest observed redshifts than at 
$z = 3$ (Stanway, {\it et al}, 2004; Bunker, A. {\it et al},
astro-ph/0403223.

We predict that this Pop III will never be found, because only a small
fraction of the baryonic matter of the universe needs to be enriched and
ionized if 99\% of the primordial baryonic matter is sequestered away in our
baryonic dark matter particles. A slow PFP evaporation, seen as
$Ly-\alpha$ clouds, would slowly enrich the universe with primordial
hydrogen/helium to maintain nearly constant elemental abundances.
Quasars probably have sufficient ultraviolet radiation to re-ionize the
universe (Escude-Miralde, Hashnelt, \& Rees, 2000). 

\section{Relationship to $Ly-\alpha$ Clouds}

The primary reason for accepting the concept of Dark Energy is the
supernova brightness curves, but in their analysis (Riess {\it et al}, 2004)
there has been no
allowance for the reduction of the transparency of the universe imposed by 
the detected Baryonic dark matter component. Significant evidence for
transmission losses might long have been seen in the quasar number density
peak at z = 1.9. 
Note that the difference between ordinary universe models and
the Goobar {\it et al}, (2002, Fig 3A) "self replenishing dust" normalized to the
supernova flux deficit at z = 0.5, is approximately 1.4 magnitudes for the
quasar number density peak at z = 1.9.

We now recognize that the existence of quasar $Ly-\alpha$ forest clouds has
evidently long been indicating the source of this transmission loss.
The presumption that such clouds must be confined by a hot intercloud
medium has been rejected on observational grounds, and the clouds should 
rapidly dissipate away unless
they are being continuously refreshed by the expected evaporation of 
our PFP objects.

It has long been known (Zuo \& Lu, 1993)
that the density of $Ly-\alpha$ clouds increases with
redshift z as $(1+ z)^{2.8}$. It has also been noticed by 
Goobar {\it et al}, (2002) that
an absorption law scaling as $(1+z)^3$ up to z = 0.5 and constant
thereafter, can explain the supernova brightness - redshift 
relationship of Riess {\it et al}, (2004). We are careful not to use the
word "grey extinction" because the transmission loss may be dominated by
refraction in the spherical lenses, which would contribute to the diffuse
background radiation of the universe, a controversial topic.

We are easily left with the following conclusion: before accepting the idea
that the Hubble expansion is dominated by a mysterious Dark Energy, it is
important to calculate the transmission to distant supernovae and quasars
as limited by the baryonic dark matter.

\section*{Acknowledgments}
I wish to thank Carl Gibson for his extended
correspondence related to the expected properties of the microlensing
particles.

\end{document}